\documentclass[Physsubmission, Phys]{SciPost}

\hypersetup{
    colorlinks,
    linkcolor={red!50!black},
    citecolor={blue!50!black},
    urlcolor={blue!80!black}
}
\usepackage{wrapfig}
\usepackage{subcaption,graphicx}
\usepackage{lineno}
\usepackage[bitstream-charter]{mathdesign}
\DeclareSymbolFont{usualmathcal}{OMS}{cmsy}{m}{n}
\DeclareSymbolFontAlphabet{\mathcal}{usualmathcal}
\begin{document}

\begin{center}{\Large \textbf{
Measurement of the $\textit{CP}$ structure of the Higgs-tau Yukawa coupling \\ 
}}\end{center}

\begin{center}
Vinaya Krishnan MB\textsuperscript{*} (on behalf of CMS collaboration)
\end{center}

\begin{center}
Institute of Physics, Bhubaneswar, India
\\
* vinaya.krishna@cern.ch
\end{center}

\begin{center}
\end{center}


\definecolor{palegray}{gray}{0.95}
\begin{center}
\colorbox{palegray}{
  \begin{tabular}{rr}
  \begin{minipage}{0.1\textwidth}
    \includegraphics[width=22mm]{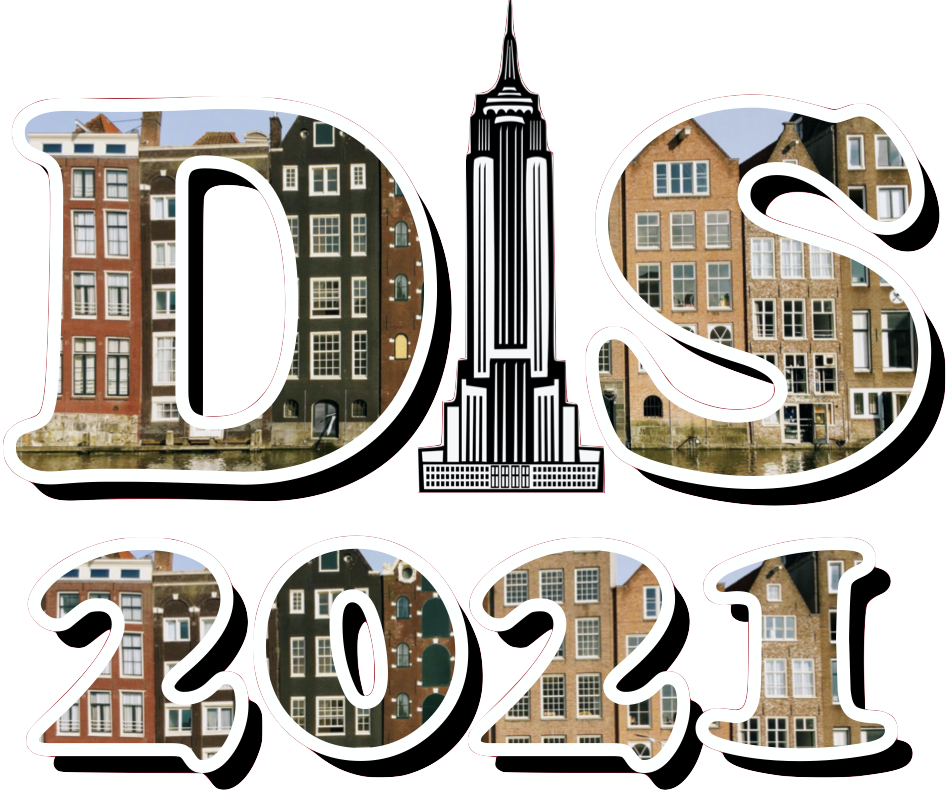}
  \end{minipage}
  &
  \begin{minipage}{0.75\textwidth}
    \begin{center}
    {\it Proceedings for the XXVIII International Workshop\\ on Deep-Inelastic Scattering and
Related Subjects,}\\
    {\it Stony Brook University, New York, USA, 12-16 April 2021} \\
    \doi{10.21468/SciPostPhysProc.?}\\
    \end{center}
  \end{minipage}
\end{tabular}
}
\end{center}

\section*{Abstract}
{\bf
The CMS experiment at the LHC has performed the first measurement of the CP structure of the Yukawa coupling between the Higgs boson and tau leptons. The measurement is based on data collected in proton-proton collisions at $\sqrt{s} = 13$ TeV during 2016-18, corresponding to an integrated luminosity of $137\text{ fb}^{-1}$. The analysis utilizes the angular correlation between the decay planes of tau leptons produced in Higgs boson decays, where dedicated analysis techniques are used to optimise the reconstruction of tau decay planes. The measured value of the CP mixing angle is $4\pm17^{\circ}$, at $68\%$ confidence level. The pure $\textit{CP-}$odd hypothesis is excluded by $3.2$ standard deviations. The analysis strategies and the results of the measurement are presented.   
}

\vspace{10pt}
\noindent\rule{\textwidth}{1pt}
\tableofcontents\thispagestyle{fancy}
\noindent\rule{\textwidth}{1pt}
\vspace{10pt}

\section{Introduction}
\label{sec:intro}
After the discovery of a Higgs boson of mass about 125 GeV at the Large Hadron Collider(\textbf{LHC})\cite{CMS:2012qbp,ATLAS:2012yve,CMS:2013btf}, many studies are being performed to ensure that the observed particle is the standard model(SM) Higgs boson. In the standard model, the Higgs boson's coupling to fermionic and bosonic fields preserves \textit{CP} symmetry, often referred to as \textit{CP}-even. The \textbf{ATLAS}  \cite{ATLAS:2017azn} and \textbf{CMS}\cite{CMS:2012vby} collaborations have already probed \textit{CP}-violating (\textit{CP}-odd) interactions of the Higgs boson to gauge bosons.  However, the \textit{CP}-odd state can couple to gauge bosons only at NLO or higher order, while its coupling to fermions can be probed at tree level.  Although these studies have excluded that the Higgs boson is a pure \textit{CP}-odd state (pseudoscalar), a \textit{CP} mixture state is not fully excluded.  

This analysis aims to access the potential mixing between a scalar and a pseudoscalar (\textit{CP}-odd state) in the Yukawa coupling to the $\tau$ leptons via the angle between tau decay planes, the analysis results discussed in the following are based on Ref\cite{CMS-PAS-HIG-20-006}. The interaction lagrangian of a Higgs boson $h$ of arbitrary \textit{CP} nature to $\tau$ leptons is described as\cite{Gritsan:2016hjl}
\begin{equation}
 \mathcal{L}_{Y} = -\frac{m_{\tau}}{v}\left(\kappa_{\tau}\bar{\tau}\tau+\bar{\kappa}_{\tau}\bar{\tau}i\gamma_{5}\tau\right)h
\end{equation}

where $m_{\tau}$ is the mass of the $\tau$ lepton, and the vacuum expectation value of Higgs field $v$ has a value of 246 GeV. The \textit{CP}-even and \textit{CP}-odd Yukawa couplings $\kappa_{\tau}$ and $\bar{\kappa}_{\tau}$ can be expressed in terms of effective mixing angle $\phi_{\tau\tau}$ as,
\begin{equation}
\tan{\phi_{\tau\tau}} = \frac{\bar{\kappa}_{\tau}}{\kappa_{\tau}} \begin{cases}
&\phi_{\tau\tau}\rightarrow 0, \textit{    CP}\text{-even}\\
&\phi_{\tau\tau}\rightarrow \frac{\pi}{2}, \textit{    CP}\text{-odd}\\
&\text{else} , \textit{    CP}\text{-mix}\\
\end{cases}
\end{equation}
We define $\phi_{\textit{CP}}$ as the angle between the $\tau$ decay planes at Higgs rest frame. This analysis measures the mixing angle ($\phi_{\tau\tau}$) from the relationship between $\phi_{\tau\tau}$ and $\phi_{\textit{CP}}$ in the differential cross-section\cite{Berge:2014sra}. 
\begin{equation}
\dfrac{d\Gamma}{d\phi_{\textit{CP}}}\propto -\cos{\left(\phi_{\textit{CP}}-2\phi_{\tau\tau}\right)}
\end{equation}
The direct access of the mixing angle from $\phi_{\textit{CP}}$ makes this analysis model-independent. The analysis is performed using full LHC Run-2 data, recorded by the CMS detector\cite{CMS:2008xjf}, corresponding to the integrated luminosity of 137 $\text{fb}^{-1}$, in the final states $\tau_{\mu}\tau_{h}$ and $\tau_h\tau_h$.
\section{$\phi_{CP}$ Reconstruction}
Tau lepton, the heaviest among the leptons, has short lifetime, and hence, decays to other lighter leptons or hadrons along with associated neutrinos. The momentum of the $\tau$-lepton is reconstructed from its decay products. However, due to the presence of neutrinos in the final state the full momentum of the $\tau$-lepton cannot be reconstructed. Therefore the decay plane is constructed from its visible decay products. The methods that are used for each decay modes are described below\cite{Berge:2014sra};
\begin{itemize}
\item \textbf{Impact parameter Method} is used for the 1-prong decays such as ($\mu^{\pm},\pi^{\pm}$), where tau decay plane is constructed from the momentum of the charged pion or hadron and its impact parameter vector. \\
\item \textbf{Neutral-pion Method} is used when tau decay products contain at least one $\pi^0$ particle. Decay planes are constructed from the momenta of charged and neutral pions. In the case of 3-prong$\left(a_{1}^{3pr}\rightarrow\pi^{\pm}\pi^{\mp}\pi^{\pm}\right)$  decay of tau lepton $\pi^{\pm}$ meson that is oppositely charged to the $a_{1}^{3pr}$ is considered as neutral pion vector for the purpose of constructing the decay plane.  
\item \textbf{Mixed Method} is used when one tau decays to one charged pion or hadron without $\pi^0$ and the other tau decays to charged prong along with the neutral pion. In this case the impact parameter method is used for the former and neutral pion method is used for the latter, respectively.   
\end{itemize}
In all these methods the decay plane is constructed in the $\pi^+\pi^-$ zero momentum frame. 
\vspace*{-\baselineskip}
\section{Analysis Strategy}
\label{sec:another}
We followed the same event selection strategy as used in the standard model Higgs to $\tau^+\tau^-$ analysis\cite{CMS-PAS-HIG-18-032}
for the $\tau_{\mu}\tau_h$ and $\tau_h\tau_h$ final state. However, we implemented some special methods to enhance the performance of this analysis.
\subsection{Special Methods}
\label{sec:another}
\textbf{Vertex Refitting:} We exclude the tracks originating from the tau decay from the vertex fitting and apply beam spot constraints to improve primary vertex resolution, which improves the impact parameter measurement\cite{Cardini:2021bar}.\\
\textbf{MVA decay mode identification:} The analysis performance is improved by utilizing a multivariate based tau decay mode identification instead of the default HPS decay modes\cite{CMS-DP-2020-041}. It enhances the assignment of the 1-$\text{ prong }+2\pi^0 (a_{1}^{1pr})$ decay mode. This provides a 20\% improvement in the expected sensitivity.
\subsection{Background Estimation}
\label{sec:another}
The main background processes to consider are: Drell-Yan ($Z/\gamma^*$), $W+\text{jets}$, $t\bar{t}$, QCD multi-jet, electroweak $W/Z$, single-top and di-boson production. All high fraction of backgrounds are estimated using data driven methods. The processes with genuine $\tau$-leptons such as $Z/\gamma^*\rightarrow\tau\tau$ and small fraction of $t\bar{t}$ and di-boson are obtained from Embedded samples. \cite{Sirunyan_2019}.
Another major background are jets misidentified as taus ($j\rightarrow\tau_h$), which is estimated using the fake factor method. \cite{CMS:2018lkr}.
The rest of the backgrounds processes like $Z/\gamma^*\rightarrow l^+l^-$ are obtained from the MC simulation.  
\subsection{Signal Extraction}
\label{sec:another}
Using a multi-classification machine learning algorithm (Neural Network for $\tau_{\mu}\tau_h$ and BDT for $\tau_h\tau_h$) events are classified into three categories. 
\begin{itemize}
\item \textbf{Higgs}: all signal processes (qqH,ggH and VH) combined into this category.
\item \textbf{Embedded}: background processes involving two genuine $\tau$-leptons.
\item \textbf{Jet-Misidentification}: background process involving at least one misidentified $\text{jet}\rightarrow\tau$-lepton fake.
\end{itemize}
The 2D unrolled $\phi_{\textit{CP}}$ distribution in the windows of increasing order of MVA score is used as the final discriminant. Due to the nature of the $\phi_{\textit{CP}}$ distribution we can exploit symmetries in the background process to reduce statistical fluctuations in MC. In the final states where impact parameter method is used to reconstruct decay plane for both the tau leptons (e.g. $\mu\pi$,$\pi\pi$), the distributions of all the backgrounds are symmetrised around the central value. 
In other final states the background distributions are flattened. However, the  $\text{jet}\rightarrow\tau$-lepton fake background distribution is symmetrised in all final states.  
\begin{figure}[!htb]
\minipage{0.50\textwidth}
\includegraphics[width=\linewidth]{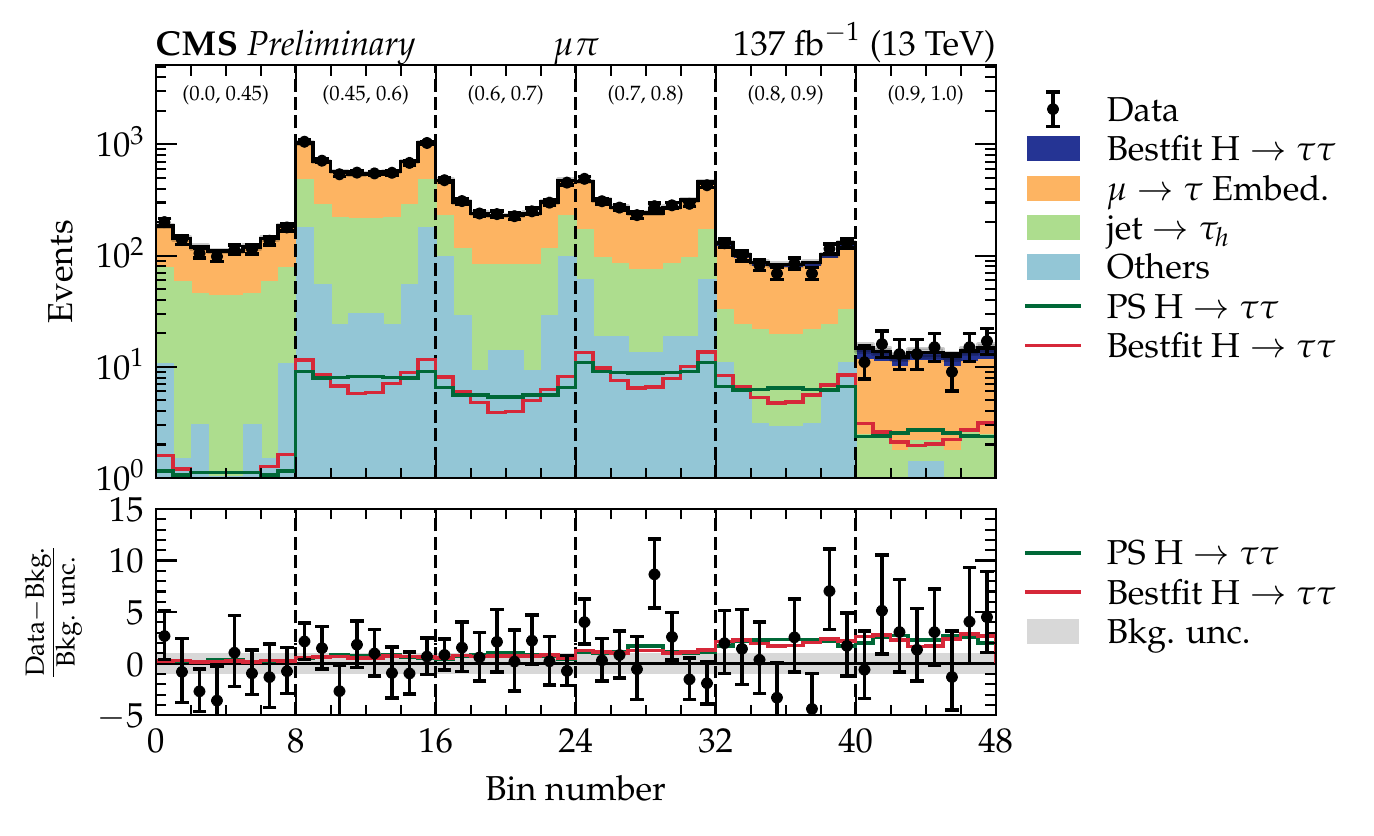}
\caption*{}
\endminipage
\minipage{0.50\textwidth}
\includegraphics[width=\linewidth]{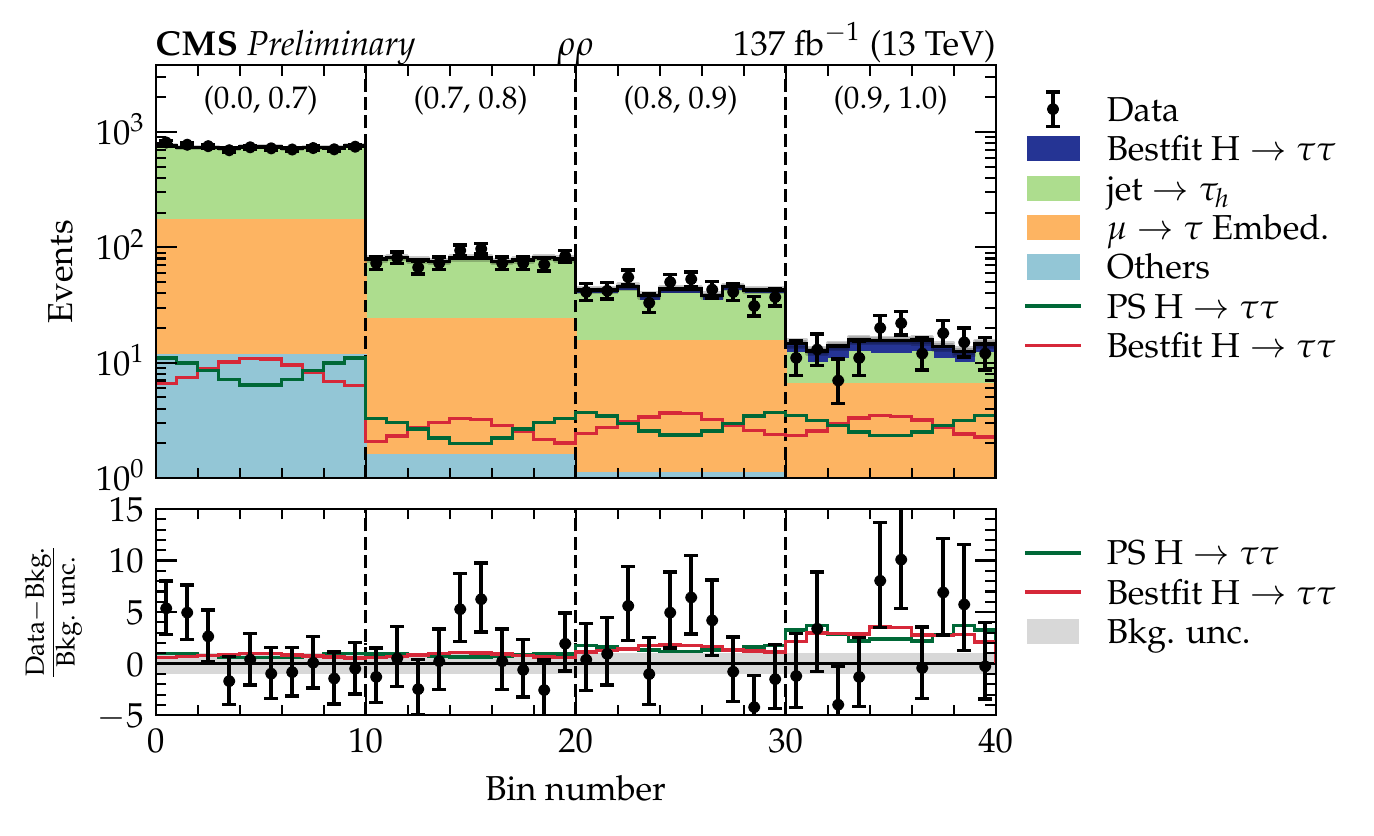}
\caption*{}
\endminipage
\vspace*{-\baselineskip}
\caption{The unrolled $\phi_{\textit{CP}}$ distributions for $\mu\pi$ and $\rho\rho$ are shown. The x-axis correspond to cyclic bins in $\phi_{\textit{CP}}$ in the range of $\left(0,2\pi\right)$. The $\mu\pi$(left) all backgrounds are symmetrised and for $\rho\rho$(right) backgrounds are flattened except $\text{jet}\rightarrow\tau$-lepton fake background\cite{CMS-PAS-HIG-20-006}}.
\vspace*{-\baselineskip}
\end{figure}\\

\section{Estimation of $\phi_{\tau\tau}$}
The estimation of $\phi_{\tau\tau}$ is obtained by the maximum likelihood fit using the enrolled $\phi_{\textit{CP}}$ distribution. The likelihood function $L(\vec{\mu},\mu^{\tau\tau},\vec{\theta})$ depends on the SM Higgs boson production signal strength ($\vec{\mu}=\mu_{ggH},\mu_{qqH},\mu_{VH}$), the $H\rightarrow\tau\tau$ decay branching fraction, \textit{CP}-mixing angle, and the nuisance parameter ($\vec{\theta}$) accounted for the systematic uncertainties.
\begin{wrapfigure}{r}{0.5\textwidth}
  \begin{center}
    \includegraphics[width=0.48\textwidth]{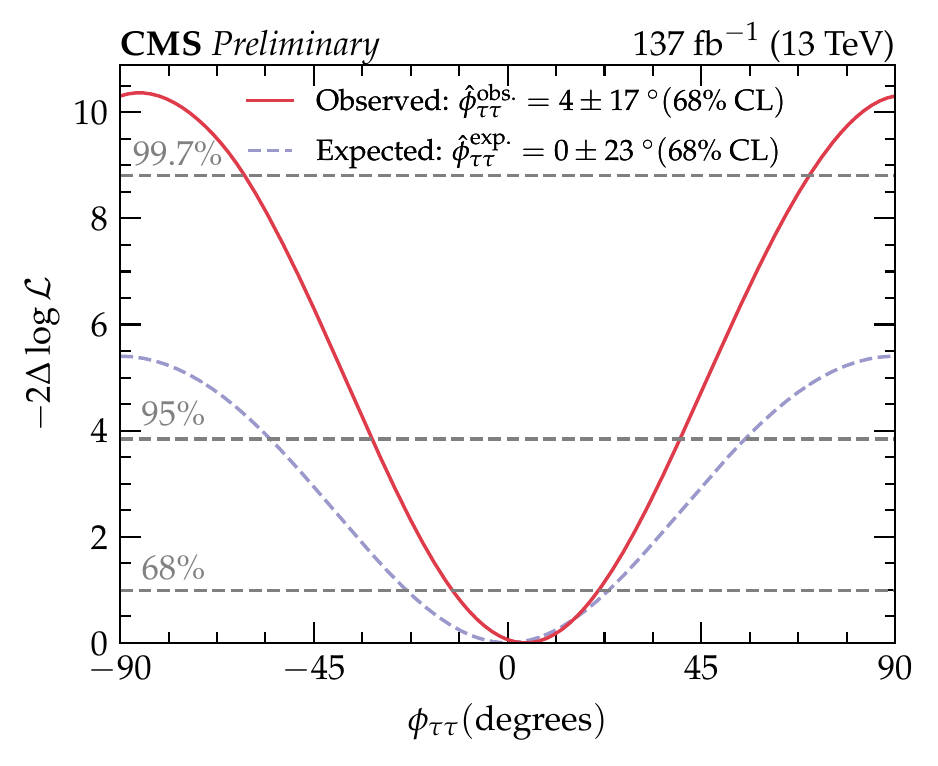}
  \end{center}
  \vspace*{-1cm}
  \caption{NLL scan on $\phi_{\tau\tau}$\cite{CMS-PAS-HIG-20-006}}
\end{wrapfigure}
 The negative log-likelihood scan for the combination (NLL) of the $\tau_{\mu}\tau_h$ and $\tau_h\tau_h$ channel shown in Figure 2,where the negative likelihood is defined as: 
\begin{equation*}
-2\Delta \ln{L} = -2\dot{\left(\ln{(L\phi_{\tau\tau})}-\ln{(L\phi_{\tau\tau}^{\text{best fit}})}\right)}
\end{equation*}
We find the 68.3, 95.5, and 99.7\% confidence intervals when $-2\Delta \ln{L}$ = 1.00,4.02 and 8.81 respectively\cite{Khachatryan_2015}. The fit favours a scalar over the pseudoscalar  $H\tau\tau$ coupling hypothesis at an observed(expected) sensitivity of 3.2(2.3) standard deviations.
The measured value of the $\phi_{\tau\tau}$ with the decomposed uncertainty\cite{CMS-PAS-HIG-20-006} is 
\begin{equation*}
\phi_{\tau\tau}=(4\pm 17(stat)\pm 2(\text{bin-by-bin})\pm 1(\text{syst})\pm 1(\text{theory}))^{\circ}
\end{equation*}
\begin{figure}[!ht]
\centering
\subcaptionbox*{}{
\includegraphics[width=5cm,height=4cm]{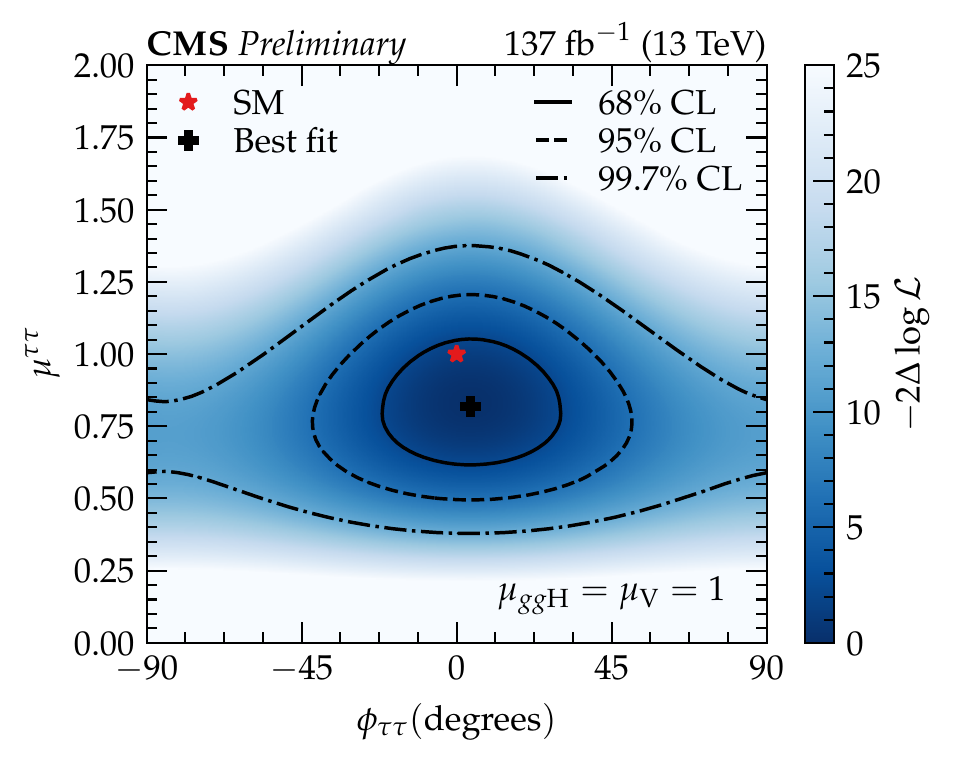}
}\qquad
\subcaptionbox*{}{
\includegraphics[width=5cm,height=4cm]{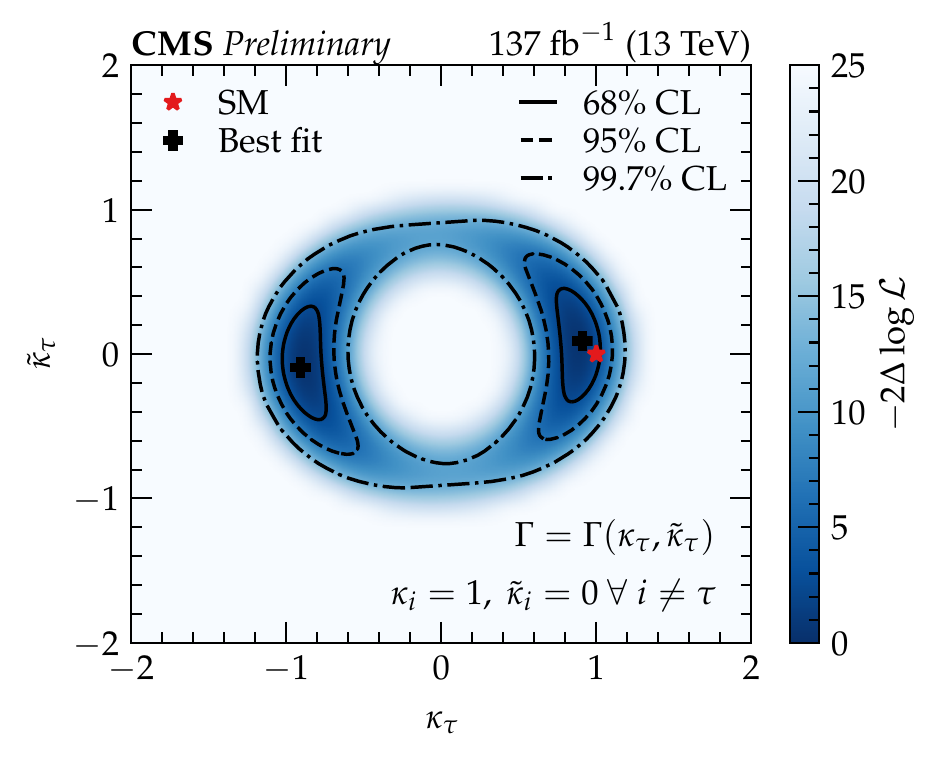}
}
\caption{2D scan of the branching fraction modifier with respect to the SM value of $\mu^{\tau\tau}$ versus $\phi_{\tau\tau}$(left). The 2D scan for scalar($\kappa$) and pseudoscalar($\bar{\kappa}$) $\tau$ Yukawa coupling (right)\cite{CMS-PAS-HIG-20-006}}
\end{figure}
 Furthermore, we performed 2D fit of the branching fraction modifier concerning the SM value $\mu^{\tau\tau}$ versus $\phi_{\tau\tau}$, where we observe that there is no strong correlation. Also, the 2D scan for scalar and pseudoscalar Yukawa coupling fit shows that the best fit value is closer to the SM prediction. 
\section{Conclusion}
A measurement is performed of the CP mixing angle $\phi_{\tau\tau}$ in the Higgs to $\tau\tau$ coupling using 137 $\text{fb}^{-1}$ of data recorded by the \textbf{CMS} experiment at centre-of-mass energy of 13 TeV. The best fit value of $\phi_{\tau\tau}$ is found to be $4\pm 17^{\circ}$. The analysis excludes a pure \textit{CP}-odd scalar at a significance of 3.2 standard deviations. The results are consistent with the standard model prediction. 
\vspace*{-\baselineskip}








\end{document}